\documentclass[10pt,journal,compsoc]{IEEEtran}

\usepackage[T1]{fontenc}

\usepackage{amsmath,amssymb,amsfonts}
\usepackage{algorithmic}
\usepackage{array}
\usepackage[caption=false,font=normalsize,labelfont=sf,textfont=sf]{subfig}
\usepackage{textcomp}
\usepackage{stfloats}
\usepackage{url}
\usepackage{verbatim}
\usepackage{graphicx}
\usepackage{cite}
\usepackage{multirow}
\usepackage{caption}

\usepackage{bm}
\usepackage{colortbl}

\usepackage{tikz}
\def\checkmark{\tikz\fill[scale=0.4](0,.35) -- (.25,0) -- (1,.7) -- (.25,.15) -- cycle;} 

\begin{document}

\title{Improving Speech Emotion Recognition with Mutual Information Regularized Generative Model}

\author{Chung-Soo Ahn, Rajib Rana, Sunil Sivadas,Carlos Busso,~\IEEEmembership{Fellow,~IEEE}, Jagath C. Rajapakse,~\IEEEmembership{Fellow,~IEEE}
\IEEEcompsocitemizethanks{
\IEEEcompsocthanksitem C-S. Ahn and J.C. Rajapakse are affiliated with the College of Computing and Data Science at Nanyang Technological University (NTU), Singapore. 
\IEEEcompsocthanksitem R. Rana is with University of Southern Queensland (USQ), Australia.
\IEEEcompsocthanksitem S. Sivadas is with NCS Group, Singapore.
\IEEEcompsocthanksitem  C. Busso is with the Language Technologies Institute, School of Computer Science, Carnegie Mellon University,  Pittsburgh, PA 15213.

Corresponding E-mail: asjagath@ntu.edu.sg}
}

\markboth{Journal of \LaTeX\ Class Files,~Vol.~14, No.~8, August~2021}%
{Shell \MakeLowercase{\textit{et al.}}: A Sample Article Using IEEEtran.cls for IEEE Journals}

\IEEEpubid{0000--0000/00\$00.00~\copyright~2021 IEEE}

\IEEEtitleabstractindextext{
\begin{abstract}
Lack of large, well- annotated emotional speech corpora continues to limit the performance and robustness of speech emotion recognition (SER), particularly as models grow more complex and the demand for multimodal systems increases. While generative data augmentation offers a promising solution, existing approaches often produce emotionally inconsistent samples due to oversimplified conditioning on categorical labels. This paper introduces a novel mutual-information-regularised generative framework that combines cross-modal alignment with feature-level synthesis.
Building on an InfoGAN-style architecture, our method first learns a semantically aligned audio–text representation space using pre-trained transformers and contrastive objectives.
A feature generator is then trained to produce emotion-aware audio features while employing mutual information as a quantitative regulariser to ensure strong dependency between generated features and their conditioning variables. We extend this approach to multimodal settings, enabling the generation of novel, paired (audio, text) features. Comprehensive evaluation on three benchmark datasets (IEMOCAP, MSP-IMPROV, MSP-Podcast) demonstrates that our framework consistently outperforms existing augmentation methods, achieving state-of-the-art performance with improvements of up to 2.6\% in unimodal SER and 3.2\% in multimodal emotion recognition. Most importantly, we demonstrate that mutual information functions as both a regulariser and a measurable metric for generative quality, offering a systematic approach to data augmentation in affective computing.

\end{abstract}

}

\maketitle

\IEEEdisplaynontitleabstractindextext
\IEEEpeerreviewmaketitle

\section{Introduction}
\label{sec:introduction}

Speech Emotion Recognition (SER) has advanced considerably over the past decade, driven by the success of deep learning architectures that can capture the complex acoustic and linguistic cues underlying human emotions. These developments have led to tremendous improvements in classification accuracy across widely used benchmark datasets \cite{Schuller2018SER, Latif2021SurveySER}. Nevertheless, compared with computer vision or natural language processing, SER research continues to face a major bottleneck: the lack of large, high-quality, and consistently annotated emotional speech corpora. This scarcity of data constrains model generalisation and makes data augmentation a crucial strategy for building reliable SER systems.

Conventional augmentation techniques, such as time-stretching, pitch-shifting, or adding Gaussian noise, expand data volume but essentially produce distorted copies of existing recordings \cite{Parthasarathy2020Ladder, zhang2017mixup}. They rarely introduce genuinely new emotional variation. In contrast, generative models can synthesise novel data by learning the underlying distribution of emotional speech. Among these, Generative Adversarial Networks (GANs) \cite{Goodfellow2014GAN} have gained particular traction due to their ability to produce realistic, high-fidelity features, and several studies have reported improved SER performance through GAN-based augmentation \cite{Yi2020AdversarialSER, Latif2021SurveySER}.

Despite their promise, most current generative approaches depend on class-conditioned GANs that assume a deterministic mapping between emotion labels and generated samples. This assumption oversimplifies human emotion, which is inherently complex and context-dependent. As a result, generated data may lack the diversity or authenticity needed for robust emotion modelling. To mitigate this issue, recent work in representation learning has explored mutual-information-based regularisation, which encourages stronger and more interpretable relationships between latent variables and generated outputs \cite{chen2016infogan, Belghazi2018MINE}. By maximising mutual information, it can be quantitatively ensured that the generated features retain meaningful emotional structure, an essential property for effective augmentation.

At the same time, emotion recognition is increasingly being treated as a multimodal task. Text and audio provide different but complementary emotional signals. While the words we choose express the meaning behind our feelings, the tone of our voice, like prosody and intonation adds a layer of expressive nuance. This duality helps convey emotions more richly and effectively \cite{Poria2017ContextSentiment, Busso_2025}. Although multimodal fusion has become common, only a few studies have explored multimodal augmentation itself. Leveraging text information to guide audio generation could substantially improve emotion modelling, yet this area remains under-explored.

Motivated by these gaps, this study introduces a mutual-information-regularised generative augmentation framework for SER. Building on the InfoGAN architecture \cite{chen2016infogan}, our approach synthesises emotion-aware speech features while preserving strong dependencies between generated features, emotion labels, and text embeddings. We further extend this concept to a multimodal setting by combining both audio and text representations, enabling the generation of new paired (audio, text) features for training emotion classifiers. The key contributions of this work are:

\begin{enumerate}
  \item  We introduce a unified adversarial learning framework for SER, which employs mutual-information regularisation to perform effective data augmentation in both unimodal (audio-only) and multimodal (audio-text) settings.
\item We uniquely design the framework for cross-modal synthesis, allowing it to leverage textual cues to generate more realistic and emotionally relevant audio features.
\item Through extensive evaluation on benchmark datasets (IEMOCAP \cite{Busso_2008_5}, MSP-IMPROV \cite{Busso_2017}, MSP-Podcast \cite{Busso_2025}), we demonstrate that our method consistently and significantly outperforms existing augmentation baselines.

\end{enumerate}

\section{Related Works}
Our research sits at the intersections of generative data augmentation for Speech Emotion Recognition (SER) and multimodal representation learning. To clarify our contribution, we will review previous work in three related areas: (1) generative models for SER, (2) multimodal and cross-modal learning for emotion recognition, and (3) strategies for addressing missing modalities. A comparative summary of this literature can be found in Table~\ref{tab:my-table}.

\subsection{ Generative Data Augmentation for SER}

The limited availability of high-quality, well-annotated emotional speech data has made data augmentation a critical component of modern SER. Traditional approaches like time-stretching, pitch-shifting, and adding noise can increase the quantity of data, but they often fail to capture the rich emotional variations necessary for accurate recognition \cite{zhang2017mixup,parthasarathy2020semi}.

To tackle these shortcomings, researchers have utilised deep generative models that can learn the complex variations in emotional speech. Generative Adversarial Networks (GANs) \cite{goodfellow2014generative} are popular in this space because they can generate high-quality features and improve the robustness of models \cite{latif2020multi,Yi2020AdversarialSER}. For example, Latif et al. \cite{latif2020multi} have introduced adversarial autoencoders specifically for SER, while Yi et al. \cite{Yi2020AdversarialSER} built upon this with specialised generators. More recently, Latif et al. \cite{latif2020augmenting} combined GANs with mix-up strategies, and Wang~et al. \cite{wang2022generative} incorporated triplet loss to tackle issues with class imbalance. 

Sahu~et al \cite{sahu2020modeling} introduced mutual information maximisation through InfoGAN \cite{chen2016infogan} to enhance emotion-controllable synthesis. Building on this idea, recent studies have explored new techniques: Latif~ et al. \cite{latif2023generative} worked on emotional text-to-speech generation, Malik~et al. \cite{malik2023preliminary} applied diffusion models for mel-spectrogram synthesis, and Kim~et al. \cite{kim2024generation} combined VAEs and diffusion techniques to boost data realism. 
Naji~et al. \cite{naji2025transformerSER} introduced a transformer-based diffusion framework that achieves near-human-level generation quality for emotional speech. Meanwhile, Roh~and~Lee \cite{roh2024conditionalDCGAN} proposed a conditional DCGAN (Deep Convolutional Generative Adversarial Network) to better control the emotional intensity in generated samples. Shilandari~et al. \cite{shilandari2024cycleganSER} also explored emotion-style transfer with CycleGANs, resulting in a noticeable improvement in emotional diversity.

\noindent{\bf Limitations and Gap:} Despite the advancements, many generative methods simplify the relationship between emotion labels and the data they generate, overlooking the complexities of emotional variability. Few studies explicitly connect the latent and generated representations. Although some use mutual information or textual conditioning \cite{sahu2020modeling,malik2023preliminary}, there is a lack of systematic mutual-information regularisation that could enhance the emotional coherence between generated audio and cross-modal signals.

\subsection{ Multimodal and Cross-Modal Learning}

Recognition of emotions is inherently a multimodal endeavor: textual semantics and acoustic prosody offer complementary insights \cite{Busso_2025}. Advances in multimodal deep learning have fostered the ability to co-learn across modalities, which improves both robustness and generalisation \cite{liang2024foundations}.

Typically, conventional methods focus on fusion or alignment of modalities \cite{goncalves2022robust,chen2023inter}, while cross-modal transfer leverages one modality to inform another. Recent research has utilised contrastive learning to strengthen shared representations \cite{liu2024contrastive} and attention-based fusion to dynamically weigh contributions from each modality \cite{goncalves2022robust}.

\noindent{\bf Limitations and Gap:} Nevertheless, these methodologies tend to be predominantly discriminative; they primarily align and fuse existing data instead of generating new instances. They optimise the representation of current data rather than synthesising novel multimodal examples. It remains a significant opportunity to exploit cross-modal dependencies not just for alignment, but for actively creating new, emotionally consistent audio-text pairs.

\subsection{Handling Missing Modalities}

Multimodal systems frequently contend with partial or noisy inputs. Researchers like Gonçalves~et al. \cite{goncalves2022robust} have employed multitask learning to maintain robustness when certain modalities are missing, while Wang~et al. \cite{wang2023exploring} explored random masking techniques. Meng~et al. \cite{meng2024deep} have even used autoencoders to reconstruct incomplete signals. More recent work by Wang~et al. \cite{wang2024incomplete} has continued to push this research agenda forward, refining methods for better recovery and analysis in multimodal scenarios.

\noindent{\bf Limitations and Gaps:} The existing methods are primarily compensatory; they focus on restoring missing modalities instead of enhancing data diversity. Their one-to-one reconstruction processes do not expand the variety of multimodal examples available for training. In contrast, our framework employs a one-to-many generative strategy, which can synthesise innovative and emotionally coherent multimodal samples.

\subsection{Our Contribution and Positioning}
Research in Speech Emotion Recognition (SER) has progressed along three parallel avenues: generative models, multimodal learning, and methods for compensating missing modalities. However, these areas have largely developed in isolation; generative models rarely incorporate cross-modal conditioning, multimodal systems are mostly discriminative, and reconstruction methods do not address data augmentation.

We integrate these strands into a mutual-information-regularised adversarial framework that:
\begin{itemize}

    \item Conducts generative augmentation in both unimodal and multimodal contexts,
    \item  Enables explicit cross-modal transfer, allowing text to inform emotion-aware audio synthesis, and
    \item  Uses mutual-information regularization to maintain a quantifiable dependency between the generated data and the conditioning signals.
\end{itemize}
This unified approach bridges the gap between discriminative multimodal learning and generative augmentation, directly addressing the challenges of data scarcity and emotional variability that continue to hinder progress in Speech Emotion Recognition.


\begin{table*}[]
\centering
\caption{Summary of literature survey}
\label{tab:my-table}
\resizebox{\linewidth}{!} {
\begin{tabular}{ccccccc}
\hline
Title                                            & Year & Generative Augmentation & Cross-modal transfer & Cross-modal alignment & Mutual Information Regularization & Multimodal augmentation \\ \hline
Latif et   al. \cite{latif2020multi}             & 2020 & \checkmark              &                      &                       &                    &                         \\\hline
Yi et   al. \cite{Yi2020AdversarialSER}               & 2020 & \checkmark              &                      &                       &                    &                         \\\hline
Latif et   al. \cite{latif2020augmenting}        & 2020 & \checkmark              &                      &                       &                    &                         \\\hline
Sahu et   al. \cite{sahu2020modeling}            & 2022 & \checkmark              &                      &                       & \checkmark         &                         \\\hline
Wang et   al. \cite{wang2022generative}          & 2022 & \checkmark              &                      &                       &                    &                         \\\hline
Latif   et al. \cite{latif2023generative}        & 2023 & \checkmark              & \checkmark           &                       &                    &                         \\\hline
Malik et   al. Model \cite{malik2023preliminary} & 2023 & \checkmark              & \checkmark           &                       &                    &                         \\\hline
Kim et   al. \cite{kim2024generation}            & 2024 & \checkmark              & \checkmark           &                       &                    &                         \\\hline
Goncalves   et al. \cite{goncalves2022robust}    & 2022 &                         & \checkmark           &                       &                    &                         \\\hline
Chen et   al. \cite{chen2023inter}               & 2023 &                         & \checkmark           & \checkmark            &                    &                         \\\hline
Wang et   al. \cite{wang2023exploring}           & 2023 &                         & \checkmark           &                       &                    &                         \\\hline
Meng et   al. \cite{meng2024deep}                & 2023 &                         & \checkmark           &                       &                    &                         \\\hline
Wang et   al. \cite{wang2024incomplete}          & 2023 & \checkmark              & \checkmark           &                       &                    &                         \\\hline
Liu et   al. \cite{liu2024contrastive}           & 2024 &                         & \checkmark           & \checkmark            &                    &                         \\ \hline
{\bf Ours}                                             & 2025 & \checkmark              & \checkmark           & \checkmark            & \checkmark         & \checkmark              \\ \hline
\end{tabular} }
\end{table*}

\section{Proposed Framework}

\begin{figure*}
    \centering
    \includegraphics[width=.7\linewidth]{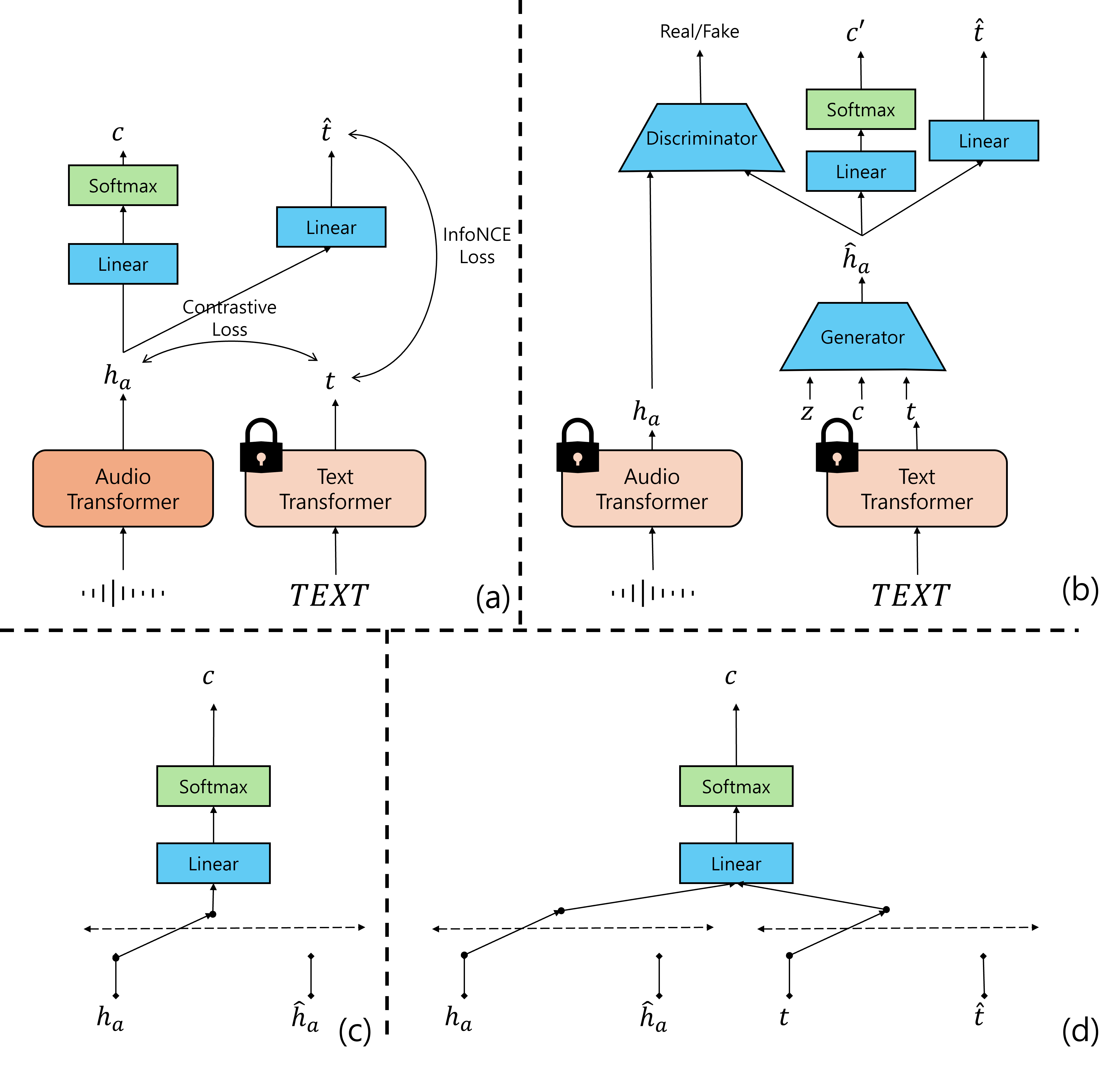}
    \caption{The visual summary of proposed augmentation framework to improve SER, consisted of three stages. First stage (a): baseline model is trained with contrastive loss and InfoNCE loss. Second stage (b): InfoGAN based feature generator is trained. Mutual information module refers to the linear layers that predict latent $c'$ and $t'$ from generated $\hat{h}$. Same weights that were trained using contrastive loss and InfoNCE loss are used in these layers as well. Third stage: we have two parallel stream in the final stage, SER case (c) and multimodal SER case (d), which is training the linear classification module with all possible input combinations by switching between $h$ or $\hat{h}$ and $t$ or $t'$.
    linear classifier layer is fine tuned with generated features.}
    \label{fig:1}
\end{figure*}

Our framework addresses the data scarcity problem in SER through a principled, three-stage approach that leverages mutual information regularization and cross-modal transfer. Rather than generating raw audio signals---a complex and often unstable process---we operate at the latent level, synthesizing discriminative embeddings directly. This strategy capitalizes on the rich representations learned by pre-trained transformers while ensuring computational efficiency and training stability. 
The framework proceeds through three core stages as depicted in Fig~\ref{fig:1}.
\begin{enumerate}
    \item Training baseline models, unimodal emotion recognition models (audio and text), which are compared against the models further trained with augmented data. The models are simultaneously trained with contrastive loss and InfoNCE loss to ensure cross-modal alignment.
    \item Learning a feature generator regularized by mutual information. During progress, mutual information loss can represent the quality of the generated sample.
    \item Augmenting the classifier with generated data. The union of generated data and training data effectively increases the size of the dataset. 
\end{enumerate}


\subsection{Baseline Model with Cross-Modal Alignment}

The first stage establishes a robust feature space. We fine-tune a pre-trained audio transformer (e.g., AST \cite{gong21b_interspeech} or Wav2Vec2 \cite{baevski2020wav2vec}) on the target SER task. The model takes a raw audio input $x_a$ and produces an audio feature vector:
\begin{align}
h &= f_a(x_a).
\end{align}
A linear classification layer with softmax activation then predicts the emotional class $c$:
\begin{align}
\hat{y} &= \text{softmax}(W_y h + b_y),
\end{align}
optimised via the cross-entropy loss:
\begin{align}
\mathcal{L}_{SER} &= -\mathbb{E}\left[\sum_{c} {1}(y=c) \log(\hat{y})\right].
\end{align}

To inject cross-modal structure and prepare for multimodal augmentation, we align the audio features with textual embeddings. Using the ground-truth transcript $x_t$, a frozen text transformer (e.g., BERT \cite{devlin2018bert}) produces a text feature:
\begin{align}
t &= f_t(x_t).
\end{align}
We then enforce a shared representation space between modalities using two complementary objectives:

\noindent{\bf Contrastive Alignment Loss ($\mathcal{L}_{CL}$):} Inspired by CLAP \cite{elizalde2023clap}, this loss pulls corresponding $(h^i, t^i)$ pairs together while pushing apart non-matching pairs within a mini-batch:
\begin{align}
\mathcal{L}_{CL} &= -\log \frac{\exp\left(\text{sim}(t^i, h^i)/\tau\right)}{\sum_{j=1}^{B} \exp\left(\text{sim}(t^i, h^j)/\tau\right)}.
\end{align}
This loss term is introduced to ensure consistency between modalities.
\noindent{\bf Mutual Information Loss ($\mathcal{L}_{MI}$):} To further maximise the dependency between $h$ and $t$, we employ an InfoNCE loss \cite{oord2018representation}. We project $h$ into the text embedding space, $\hat{t} = W_t h + b_t$, and compute:
\begin{align}
\mathcal{L}_{MI} &= -\log \frac{\exp\left(\text{sim}(t^i, \hat{t}^i)/\tau\right)}{\sum_{j=1}^{B} \exp\left(\text{sim}(t^i, \hat{t}^j)/\tau\right)}.
\end{align}
Contrastive alignment loss aims to enhance similarity between $h$ and $t$. On the other hand, mutual information loss aims to recover $t$ solely from $h$, which is identically maximizing similarity between $t$ and $\hat{t}$.

The baseline model is jointly optimized to minimize:
\begin{align}
\mathcal{L}_{Baseline} &= \mathcal{L}_{SER} + \alpha \mathcal{L}_{CL} + \beta \mathcal{L}_{MI},
\end{align}
where $\alpha$ and $\beta$ are weighting hyperparameters. This results in an audio feature space that is not only discriminative for emotion but also semantically aligned with textual descriptions.

\subsection{ Mutual Information Regularised Feature Generation}

The second stage learns a generative model to synthesize new audio features from the distribution of the aligned feature space. We build upon the InfoGAN architecture \cite{chen2016infogan}, which augments the standard GAN objective with a mutual information term to ensure the generated features are interpretable and controllable.

Our framework instantiates a generator $G$ and a discriminator $D$. The generator takes a noise vector $z \sim p_z(z)$, an emotion label $c$, and a text embedding $t$ to produce a synthetic audio feature:
\begin{align}
\hat{h} &= G(z, c, t).
\end{align}
The discriminator $D$ is trained to distinguish between real features $h$ and generated features $\hat{h}$. The adversarial objective follows the standard formulation:
\begin{align} \begin{split}  
\min_G \max_D V(D, G) &= \mathbb{E}_{h \sim p{\text{data}}(h)}[\log D(h)]\\ 
& \quad \quad + \mathbb{E}_{z, c, t}[\log(1 - D(\hat{h}))].
\end{split} \end{align}

The key to our approach is ensuring that the generated feature $\hat{h}$ retains strong dependencies on the conditioning variables $c$ and $t$. We achieve this goal by maximizing the mutual information $I((c, t); \hat{h})$. In practice, this is implemented by adding auxiliary prediction losses that force the generator to produce features from which $c$ and $t$ can be reconstructed.

We re-use the pre-trained prediction layers from the baseline model ($W_y, b_y$ and $W_t, b_t$), freezing them during this stage. This weight-sharing strategy ensures the latent codes $c$ and $t$ are forced to represent emotional and textual information, respectively.

The mutual information losses are defined as:

\noindent{\bf Emotion Prediction Loss:} A cross-entropy loss ensuring the generated feature $\hat{h}$ predicts the correct emotion label $c$:
\begin{align}
\hat{y}_g &= \text{softmax}(W_y \hat{h} + b_y), \\
\mathcal{L}_{I_y} &= -\mathbb{E}\left[\sum_{c} {1}(y=c) \log(\hat{y_g})\right].
\end{align}
\noindent{\bf Text Prediction Loss:} An InfoNCE loss that attracts the projected text embedding from $\hat{h}$ to the true conditioning text embedding $t$:
\begin{align}
\hat{t_g} &= W_t \hat{h} + b_t, \\ 
\mathcal{L}_{I_t} &= -\log \frac{\exp\left(\text{sim}(t^i, \hat{t_g}^i)/\tau\right)}{\sum_{j=1}^{B} \exp\left(\text{sim}(t^i, \hat{t_g}^j)/\tau\right)}.
\end{align}
The full objective for the InfoGAN is:
\begin{align}
\min_{G}\max_{D} V(D, G) - \lambda_y \mathcal{L}_{I_y} - \lambda_t \mathcal{L}_{I_t},
\end{align}
where $\lambda_y$ and $\lambda_t$ are hyperparameters controlling the strength of the mutual information regularisation.

\subsection{Data Augmentation and Classifier Fine-tuning}

In the final stage, we leverage the trained generator for data augmentation. All feature extraction and generation modules are frozen, and only the classification layers are fine-tuned.

\noindent{\bf Unimodal SER:} We generate a set of synthetic audio features ${\hat{h}}$ equal in size to the original training set. The classifier is then fine-tuned on the combined set ${h} \cup {\hat{h}}$, effectively doubling the training data.

\noindent{\bf Multimodal SER:} We generate paired sets ${\hat{h}}$ and ${\hat{t}}$. A new classifier that takes concatenated audio-text features is then trained on all possible combinations of real and generated features: $(h, t), (h, \hat{t}), (\hat{h}, t), (\hat{h}, \hat{t})$. This strategy quadruples the number of multimodal training instances, providing a substantial boost in data diversity and volume.

This staged approach ensures that the augmented data consists of novel, emotionally coherent, and cross-modally consistent features that directly address the core challenges of data scarcity in SER.

\section{Experiment}
\subsection{Dataset}
\subsubsection{IEMOCAP (Interactive Emotional Dyadic Motion Capture) dataset \cite{Busso_2008_5}}

The IEMOCAP dataset is a standard benchmark for SER research, which consists of approximately 12 hours of audiovisual recordings, featuring dyadic interactions between 5 pairs of male and female actors. Multimodal raw data was collected and we used speech recording and transcript for our experiments. The sessions were designed to elicit emotional expressions through scripted and improvised dialogues. Each recording was segmented per utterance and manually annotated by multiple human evaluators. Although nine categorical emotion labels are provided, four emotions are considered: neutral, happiness (merged with excited), sadness, and anger.

\subsubsection{MSP-IMPROV (Multimodal Signal Processing - Interactive Emotion Dyadic Motion Capture) dataset \cite{Busso_2017}}

The MSP-IMPROV dataset is a multimodal corpus for speech emotion recognition (SER) research, with a structure similar to that of IEMOCAP. It comprises approximately 9 hours of recordings from 12 actors (6 male and 6 female) interacting with each other. The data samples are segmented at the utterance level and annotated by multiple human evaluators. The primary categorical emotion labels are of four classes: happiness, sadness, anger, and neutral.

\subsubsection{MSP-Podcast \cite{Busso_2025}}

MSP-Podcast corpus is constructed from English speech collected from existing podcast recordings. It is not recorded in a controlled laboratory environment and thus more closely resembles an in-the-wild speech dataset. The dataset is provided with predefined training, development, and test partitions. The MSP-PODCAST dataset includes a larger set of emotion categories than MSP-IMPROV; however, to maintain consistency with our experimental setup, only four categories are considered: neutral, happiness, sadness, and anger.

\subsection{Model configuration}
\subsubsection{Data augmentation for SER}



We adopt the pre-trained AST (Audio Spectrogram Transformer) \cite{gong21b_interspeech} from `https://huggingface.co/MIT/ast-finetuned-audioset-10-10-0.4593' as a backbone. And add linear classifier after pooled output from AST. For text feature encoder, we use the pre-trained BERT \cite{devlin2018bert}, that is 'bert-uncased-base' version. For SER experiment, we do not consider multimodal SER accuracy, thus, BERT is frozen and we only use it to extract text features from ground-truth transcript. We intended to validate that our data augmentation framework is better than existing methods. So feature encoders in this experiment were configured by relatively smaller models. 


We train InfoGAN to generate the feature encoding, which is in the dimension of 768. So we take the simplest form of architecture for Discriminator and Generator which is a linear layer without any activation function. Also the network for mutual information maximization is also linear layer only. Especially, the linear layer that maximizes mutual information with \(y\) uses the same layer from emotion classifier. In this way, we can prevent latent code vector to encode something other than emotion, as there are many other complex variations inherent in the training dataset. Emotional code vector is trained with cross entropy to maximize the mutual information. For the latent embedding given from BERT embedding we use the InfoNCE loss, which is a contrastive learning loss trying to attract the BERT embedding and predicted text embedding from generated sample. Lastly, we adopted a mix-up strategy for training the generator and the discriminator as proposed in \cite{zhang2017mixup}, for stable training of the generator.

\subsubsection{Data augmentation for multimodal SER}


As multimodal data augmentation method is our key contribution, we selected feature encoders that perform close to state-of-the-art. We adopt the pre-trained Wav2Vec2 \cite{10089511} from `https://huggingface.co/audeering/wav2vec2-large-robust-12-ft-emotion-msp-dim' for multimodal SER experiment. For the text feature encoder, we used the pre-trained RoBERTA \cite{devlin2018bert}, that is 'roberta-uncased-large' version. For multimodal SER experiment, RoBERTa is separately trained to predict emotion labels from transcript before the baseline stage. After that, the RoBERTa model weights are frozen during training baseline and onwards. For the MSP-Podcast experiment, we use Llama 2.0 \cite{touvron2023llama} from `https://huggingface.co/meta-llama/Llama-2-7b' instead of RoBERTa for text features, but the same Wav2Vec2 for audio features.
Other than the change of feature size, due to change of back-bone model, the configuration for the InfoGAN stay the same. Multimodal SER classification layer takes concatenated feature of both modalities. The classifier consists linear layer,  one hidden layer (256) or two hidden layers (256-128), for IEMOCAP, MSP-IMPROV or MSP-Podcast respectively.

\subsubsection{Hyperparameters}

We used Adam optimizer with a $1e-4$ learning rate for training. Different weights were applied for each class during cross entropy loss computation due to class imbalance, for the MSP-IMPROV and MSP-Podcast. The weights were inverse proportional to the total numbers of samples per each class. LoRA (Low-Rank Adaptation of Large Language Model)\cite{hu2022lora} was adopted when Llama was trained for text emotion recognition. The weighting factors for each loss terms ($\alpha$, $\beta$, $\lambda_{y}$ and $\lambda_{t}$) were fixed to $1$.



\section{Result}

The experiment was conducted in two rounds. First, our data augmentation framework was tested on SER case. We observe the performance difference before and after data augmentation. We conduct an ablation study to check the contributing effects of each modules in the model. Second, we run the multimodal SER experiments. 

\subsection{SER experiment}

We experimented our method on the commonly used dataset IEMOCAP. The experiment followed leave-one-speaker-out cross validation scheme. Unweighted average recall is reported per each speaker. This experiment compares the effect of different data augmentation methods in SER. Thus, we followed the same experiment protocols as in previous work in this context (we only used the IEMOCAP dataset). We statistically average and present mean and standard deviation in Table \ref{tab:my-table1}. 
\begin{table}[]
\centering
\caption{Performance comparison with existing data augmentation methods on IEMOCAP dataset}
\label{tab:my-table1}
\resizebox{\columnwidth}{!}{%
\begin{tabular}{ccc} \hline
Methods               & Without augmentation & With augmentation \\ \hline
Sahu   et al. \cite{sahu2020modeling}         & 59.42                & 60.29             \\
Bao   et al. \cite{bao2019cyclegan}         & 59.48±0.71           & 60.37±0.70        \\
Latif   et al. \cite{latif2022self}       & { 60.51±0.57}           & 61.05±0.68        \\
Malik   et al. \cite{malik2023preliminary} & 58.62±2.11           & 61.22±1.85        \\
 
{ Ours}                  & {\bf 60.81±4.83}           & {\bf 63.40±2.52}       \\\hline
\end{tabular}%
}
\end{table}

The results depict that our augmentation method achieves the state-of-the-art performance among existing methods. Furthermore, our augmentation method shows highest improvement against its baseline after augmentation, which is $2.6\%$, which is the same amount of improvement obtained by Malik et al. \cite{malik2023preliminary}. Sahu et al. employed typical emotional label conditional GAN and Bao et al. \cite{bao2019cyclegan} CycleGAN approach to generate additional data to train SER model. Latif et al. \cite{latif2022self} achieved generalization improvement by using unlabelled dataset during training. Still, the improvement of accuracy against their baseline was marginal. Recently, the margin between baseline and data augmented SER was improved when data generation model was conditioned on text input as in Malik et al. \cite{malik2023preliminary}. Our model also achieved the same margin and state-of-the-art performance by using mutual information between generated data and text data. From the perspective of range (from mean minus standard deviation to mean plus standard deviation), we observe that the lower end of the range from our model is higher than others, depicting that our model can consistently outperform existing models.

\subsubsection{Ablation study}

We performed ablation study to examine the effect of each modules we proposed. The results are depicted in Table \ref{tab:my-table2}.

\begin{table}[]
\centering
\caption{Experiment results from ablation study}
\label{tab:my-table2}
\resizebox{\columnwidth}{!}{%
\begin{tabular}{cc}
 \hline
Methods                                   & UAR        \\\hline
 
Full   model                              & 63.40±2.52 \\

Without cross-modal alignment & 62.31±3.65 \\
 
Without  cross-modal alignment \& text embedding               & 61.07±2.45 \\
 
Without  cross-modal alignment \& mutual information maximization & 61.70±2.58 \\

Without cross-modal alignment\& data augmentation & 60.73±3.92 \\
\hline
\end{tabular}%
}
\end{table}

First, we present the results when cross-modal alignment component(text embedding component and corresponding losses) of the first stage is removed. This result shows that the cross-modal alignment module during the first stage improves the quality of data augmentation. We hypothesize that the loss terms corresponding to cross-modal alignment enforces strong dependence between audio embedding and text embedding, providing better initial state to train InfoGAN. Next, we observed the contribution of text embedding and mutual information maximization loss in InfoGAN training. The result depicts that both components has significant contribution in quality of data augmentation. Lastly, we experiment the case where we do not use data augmentation and cross-modal alignment being removed in the first stage, which is the SER performance on fine-tuning on vanilla AST. The result shows similar performance to that, before data augmentation, as in Table \ref{tab:my-table2}. This implies that cross-modal alignment helps improve the quality of generated samples from InfoGAN, not directly improving SER performances.

\subsection{Multimodal SER Experiment}

Multimodal SER experiment was performed on three datasets. This experiment is an use-case-scenario, where you are provided with two emotion prediction models, that one takes speech input and text input for the other. Through this experiment, we intend to show that even with simple data augmentation method via generative model, we improve the performances to state-of-the-art. In this experiment, the previous works \cite{poria2017context, zhao2021missing, liu2024contrastive} did not report the standard deviation, thus, we do not present standard deviation for comparison.

\subsubsection{IEMOCAP}

\begin{table}[]
\centering
\caption{Multimodal SER experiments on IEMOCAP dataset}
\label{tab:my-tableadsfads}
\begin{tabular}{cccc}
\hline
\multirow{2}{*}{Model} & \multicolumn{3}{c}{Unweighted Accuracy}                            \\
                       & Audio                & Text                 & Audio + Text         \\ \hline
sc\_LSTM \cite{poria2017context}               & 55.96                & 63.51                & 68.45                \\
bc\_LSTM \cite{poria2017context}               & 57.43                & 67.69                & 73.89                \\
AE \cite{bengio2006greedy}                    & 56.79                & 25.44                & 26.25                \\
CRA \cite{tran2017missing}                   & 56.79                & 27.96                & 30.13                \\
MMIN \cite{zhao2021missing}                  & 58.23                & 67.52                & 74.72                \\
CIF-MMIN \cite{liu2024contrastive}              & 58.44                & {69.26}                & 75.65                \\
{ Ours}                   & {72.83}                &{  65.51}                & {\bf 76.54}                \\ \hline
\multicolumn{1}{l}{}   & \multicolumn{1}{l}{} & \multicolumn{1}{l}{} & \multicolumn{1}{l}{} \\
\multicolumn{1}{l}{}   & \multicolumn{1}{l}{} & \multicolumn{1}{l}{} & \multicolumn{1}{l}{} \\
\multicolumn{1}{l}{}   & \multicolumn{1}{l}{} & \multicolumn{1}{l}{} & \multicolumn{1}{l}{} \\
\multicolumn{1}{l}{}   & \multicolumn{1}{l}{} & \multicolumn{1}{l}{} & \multicolumn{1}{l}{}
\end{tabular}
\end{table}

The experiment result revealed the advantage of our data augmentation method, in terms of accuracy. CRA \cite{tran2017missing}, MMIN \cite{zhao2021missing} and CIF-MMIN \cite{liu2024contrastive} exploit the dependency between features from different modality, thus achieving good performance among other existing works. Inspired from these results, we adopted similar methodology, not to recover missing feature (modality) but to generate new multimodal feature sets. Our text emotion recognition model was rather naive implementation which shows not the best performance, yet, our simple multimodal data augmentation achieved state-of-the-art. 

\subsubsection{MSP-IMPROV}

\begin{table}[]
\centering
\caption{Multimodal SER experiments on MSP-IMPROV dataset}
\label{tab:my-tablejdkffl}
\begin{tabular}{cccc}
\hline
\multirow{2}{*}{Model} & \multicolumn{3}{c}{Unweighted Accuracy}                            \\
                       & Audio                & Text                 & Audio + Text         \\ \hline
sc\_LSTM \cite{poria2017context}                & 40.91                & 32.01                & 41.36                \\
bc\_LSTM \cite{poria2017context}                & 42.53                & 57.39                & 58.32                \\
AE \cite{bengio2006greedy}                    & 42.69                & 27.62                & 38.73                \\
CRA \cite{tran2017missing}                    & 38.96                & 28.37                & 37.97                \\
MMIN \cite{zhao2021missing}                   & 42.71                & 56.49                & 60.98                \\
CIF-MMIN \cite{liu2024contrastive}              & 41.56                & {58.57}                & 61.72                \\
{ Ours}                   & {57.85}                & { 40.29}                & {\bf 62.84}                \\ \hline
\multicolumn{1}{l}{}   & \multicolumn{1}{l}{} & \multicolumn{1}{l}{} & \multicolumn{1}{l}{} \\
\multicolumn{1}{l}{}   & \multicolumn{1}{l}{} & \multicolumn{1}{l}{} & \multicolumn{1}{l}{} \\
\multicolumn{1}{l}{}   & \multicolumn{1}{l}{} & \multicolumn{1}{l}{} & \multicolumn{1}{l}{} \\
\multicolumn{1}{l}{}   & \multicolumn{1}{l}{} & \multicolumn{1}{l}{} & \multicolumn{1}{l}{}
\end{tabular}
\end{table}

We ran the same experiment with the MSP-IMPROV dataset. As IEMOCAP and MSP-IMPROV are datasets with similar characteristics, similar trend with IEMOCAP experiment result is observed (refer to Table.\ref{tab:my-tablejdkffl}). 

\subsubsection{MSP-Podcast}

\begin{table}[]
\centering
\caption{Multimodal SER experiments on MSP-Podcast dataset}
\label{tab:my-table3}
\begin{tabular}{ll}
\hline
\multicolumn{1}{c}{Model}                                  & \multicolumn{1}{c}{Unweighted Accuracy} \\ \hline
\multicolumn{1}{c}{emoDARTS \cite{rajapakshe2024emodarts}} & \multicolumn{1}{c}{61.15 ± 2.41}        \\
\multicolumn{1}{c}{Ours}                                   & \multicolumn{1}{c}{60.00 ± \bf {0.72}}        \\ \hline
                                                           &                                         \\
                                                           &                                         \\
                                                           &                                        
\end{tabular}
\end{table}

Unlike, IEMOCAP and MSP-IMPROV, MSP-Podcast is more naturalistic dataset with much larger number of samples. As there weren't any existing works that ran experiment with similar protocol, we compared our method with emoDARTS \cite{rajapakshe2024emodarts}, which uses neural architecture search and only audio input (refer to Table.\ref{tab:my-table3}). While the performance on both methods are similar, we note that our method is more stable as the standard deviation across trials is smaller. 
Furthermore, compared to emoDARTS, our model offer a simpler architecture.

Notable point is, that MSP-Podcast has severe class imbalance problem, thus, we generated data samples for the lesser count of emotional classes to obtain the same numbers of samples per emotion. This scheme of data generation is unlike the generation in existing works, as \cite{wang2024incomplete}, which requires other modality features to transfer emotional information. In this case, we generated audio feature and text feature from scratch, with only emotional label provided. Still, our data augmentation method synthesized good sample and we were able to mitigate the class-imbalance problem.  

\subsection{Qualitative analysis of generated features}
\begin{figure*}[h]
    \centering
    \captionsetup{justification=centering}
    \includegraphics[width=\linewidth]{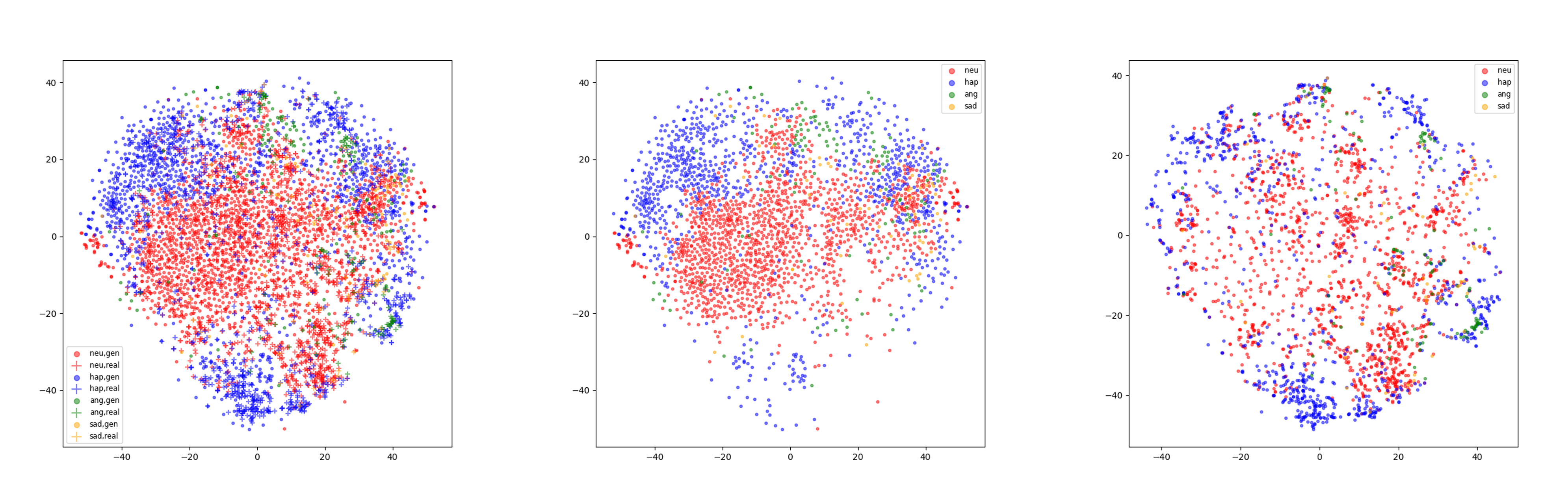}
    \caption{T-SNE plot for both [real and generated combined] (left), {\bf generated} (center) and real (right) audio features}
    \label{fig:2}
\end{figure*}
We visualized both generated features from InfoGAN and real audio features for qualitative analysis. The data samples from the test dataset of the MSP-Podcast was used and same number of generated features were prepared. Scatter plots of the features, after transformation via t-sne \cite{maaten2008visualizing}, were produced and analysed. It is clearly observable that features from different classes are placed on different separable regions (refer to Fig.\ref{fig:2}).



 
 
 
 

\section{Conclusion}
We have presented a mutual-information-regularised generative framework that effectively addresses data scarcity in speech emotion recognition through principled cross-modal augmentation. Unlike conventional approaches that rely on simplistic label conditioning, our method establishes quantifiable dependencies between generated features and rich multimodal signals, ensuring the emotional coherence and diversity of synthesised data.

The key insights from our work are threefold. First, mutual information regularisation provides both a training objective and a measurable quality metric, enabling controllable generation of emotionally consistent features. Second, cross-modal transfer from text to audio allows for more nuanced emotion modeling, capturing the complex interplay between linguistic content and vocal expression. Third, our unified framework seamlessly extends to multimodal augmentation, generating paired audio-text features that significantly enhance multimodal SER performance.

Extensive experiments across three diverse datasets validate our approach, demonstrating consistent improvements over existing methods in both unimodal and multimodal settings. The ablation studies further confirm the individual contributions of cross-modal alignment and mutual information regularization to the overall framework's effectiveness.


Despite these promising results, the current work has limitations. The framework assumes access to aligned speech–text pairs (for obtaining embeddings) and relies on a pretrained SER encoder, which may not be available for all languages or domains. Moreover, the GAN synthesises latent representations rather than raw waveforms, so the realism of generated speech may be limited. Future work will address these issues by exploring end-to-end emotional speech synthesis (e.g. using text-to-speech or diffusion-based generative models) to produce waveform-level samples, potentially improving audio fidelity. We also plan to investigate unsupervised or weakly supervised extensions that reduce reliance on transcripts, and to apply the method to multilingual or continuous-emotion settings. These directions may further advance data-efficient SER by leveraging generative models to capture the nuance of emotional speech.

\bibliographystyle{IEEEtran}
\bibliography{refs}

\end{document}